\documentclass[english]{article}
\usepackage[T1]{fontenc}
\usepackage[latin9]{inputenc}
\usepackage{float}
\usepackage{amsmath}
\usepackage{amssymb}
\usepackage{graphicx}

\makeatletter
\usepackage{jheppub} 

\author{Piotr Kucharski} 

\affiliation{Walter Burke Institute for Theoretical 
Physics, California Institute of Technology, 
Pasadena,~CA~91125,~USA} 
\affiliation{Faculty of Physics, University of Warsaw, 
ul. Pasteura 5, 02-093 Warsaw, Poland}
\emailAdd{piotrek@caltech.edu}
\abstract{
$\hat{Z}$ invariants of 3-manifolds were introduced as 
 series in $q=e^{2\pi i\tau}$ in order to categorify 
Witten-Reshetikhin-Turaev invariants corresponding to 
$\tau=1/k$. However modularity properties suggest that
all roots of unity are on the same footing. 
The main result of this paper is the expression 
connecting  Reshetikhin-Turaev invariants with $\hat{Z}$
 invariants for $\tau\in\mathbb{Q}$. We present the 
reasoning leading to this conjecture and test it on 
various 3-manifolds.
}

\makeatother

\usepackage{babel}
\begin{document}
\title{$\hat{Z}$ invariants at rational $\tau$}

\maketitle
\newpage{}

\section{Introduction and summary}

The main goal of this paper is to explore the behaviour of $\hat{Z}$
invariants of 3-manifolds at rational $\tau$ (in general $\tau\in\mathbb{H}$
-- the upper half-plane). $\hat{Z}$ invariants were introduced in
\cite{GPV1602,GMP1605,GPPV1701,CCFGH1809} as series in $q=e^{2\pi i\tau}$
with integer coefficients in order to enable the categorification
of Witten-Reshetikhin-Turaev (WRT) invariants of 3-manifolds. It turns
out that, apart from the topological applications, $\hat{Z}$ invariants
are very interesting from the point of view of physics and number
theory.

Physically $\hat{Z}$ invariant is a 3d analogue of the elliptic genus
introduced in \cite{Wit87}. More precisely it is a\,supersymmetrix
index of 3d $\mathcal{N}=2$ theory with 2d $\mathcal{N}=(0,2)$ boundary
condition studied first in\,\cite{GGP1302}. Detailed analysis of
this interpretation can be found in\,\cite{GPV1602,GPPV1701}, whereas
\cite{Chung1811} provides a lot of explicit results for various examples.
$\hat{Z}$ invariants are also related to 2d logarithmic conformal
field theories\,\cite{CCFGH1809} and newly proposed two-variable
series for knot complements\,\cite{GM1904}.

Due to their modular properties, $\hat{Z}$ invariants are interesting
from the point of view of number theory. A broad discussion of this
subject can be found in \cite{CCFGH1809}. For us the most important
are two aspects. Firstly, for many 3-manifolds $\hat{Z}$ invariants
can be expressed as a linear combination of false theta functions
\cite{GMP1605,CCFGH1809,Chun1701}. This fact plays an important role
in explicit calculations in Sections \ref{subsec:Brieskorn-spheres}
and\,\ref{subsec:Other-Seifert-manifolds}. An analogous property
for WRT invariants was studied earlier in \cite{LZ99,Hik0405,Hik0409,Hik0504,Hik0506,Hik0604,Hik11}.

In order to understand the second aspect, let us make a step back
to the relation between WRT\,invariants and $\hat{Z}$\,invariants
for plumbed 3-manifolds \cite{GPV1602,GMP1605,GPPV1701,CCFGH1809}
\begin{align}
\textrm{WRT}[M_{3}(\Gamma);1/k]= & \underset{q\rightarrow e^{\frac{2\pi i}{k}}}{\lim}\frac{\sum_{a\in\textrm{Coker}M}e^{-2\pi ik(a,M^{-1}a)}\sum_{b\in2\textrm{Coker}M+\delta}S_{ab}\hat{Z}_{b}}{2\left(q^{1/2}-q^{-1/2}\right)},\nonumber \\
S_{ab}= & \frac{e^{-2\pi i(a,M^{-1}b)}}{|\det M|^{1/2}},\label{eq:WRT from Zhat intro}
\end{align}
where $M$ is the linking matrix of the plumbing graph $\Gamma$ (for
details see Section\,\ref{subsec:Plumbed-3-manifolds}). Equation
(\ref{eq:WRT from Zhat intro}) corresponds to $\tau=1/k$. In this
case there exists a well-known physical interpretation in the language
of Chern-Simons theory, where $k\in\mathbb{N}$ is the quantum-corrected
Chern-Simons level \cite{Witten_Jones} (in the whole paper we restrict
to the $SU(2)$ gauge group). However from the point of view of number
theory $\tau=1/k$ is conceptually on the same footing as all other
rational numbers \cite{LZ99}. Therefore there arises a natural question
(which is the main motivation of this work):
\noindent \begin{center}
What happens with (\ref{eq:WRT from Zhat intro}) for $\tau=r/s$?
\par\end{center}

Since for $\tau=r/s$ ($r,s\in\mathbb{Z}$) there is no Chern-Simons
theory interpretation, we will refer to the left hand side as the
Reshetikhin-Turaev (RT) invariant -- their combinatorial definition
using quantum group representation theory \cite{RT91} works for all
$\tau\in\mathbb{Q}$. The main result of this paper is the following
expression connecting the RT invariant with the $\hat{Z}$ invariant
\begin{align}
\textrm{RT}[M_{3}(\Gamma);r/s]= & \underset{q\rightarrow e^{2\pi i\frac{r}{s}}}{\lim}\frac{\sum_{a\in\textrm{Coker}(rM)}e^{-2\pi i\frac{s}{r}(a,M^{-1}a)}\sum_{b\in2\textrm{Coker}M+\delta}S_{ab}\hat{Z}_{b}}{2\left(q^{1/2}-q^{-1/2}\right)G(s,r)^{L}},\nonumber \\
S_{ab}= & \frac{e^{-2\pi i(a,M^{-1}b)}}{|\det M|^{1/2}},\label{eq:RT from Zhat intro}\\
G(s,r)= & \sum_{c\in\mathbb{Z}_{r}}e^{2\pi i\frac{s}{r}c^{2}},\nonumber 
\end{align}
where values of the quadratic Gauss sum are discussed in Section \ref{sec:Main-conjecture}.
We checked this formula in many examples and conjecture that it is
true for all plumbed 3-manifolds. We expect that similar formula holds
for all 3-manifolds, but in that situation obtaining $\tau\rightarrow r/s$
limit of $\hat{Z}$ and testing is problematic.

The form of (\ref{eq:RT from Zhat intro}), especially the summation
over $a\in\textrm{Coker}(rM)$, is quite surprising. Is $M\mapsto rM$
a purely computational phenomenon or does it have a topological interpretation?
If the latter is true, should we view $rM$ as the matrix defining
a 3-manifold? What would be the relation to the initial one? We will
come back to these questions in Sections \ref{sec:Main-conjecture}
and \ref{sec:Open-questions}.

The plan of this paper is as follows. Section \ref{sec:Prerequisites}
contains the necessary preparations, focusing on plumbed 3-manifolds
and an expression for RT invariant independent of (\ref{eq:RT from Zhat intro}).
In Section \ref{sec:Main-conjecture} we derive and discuss our main
result -- the formula (\ref{eq:RT from Zhat intro}). Tests on various
examples are presented in Section \ref{sec:Examples}. Finally, Section
\ref{sec:Open-questions} is devoted to the future directions.

\emph{Remark: }Soon after this paper appeared on arXiv, an independent
approach to $\hat{Z}$ invariants at rational $\tau$ was presented
in \cite{Chung1906}.

\section{Prerequisites \label{sec:Prerequisites}}

\subsection{Plumbed 3-manifolds\label{subsec:Plumbed-3-manifolds}}

\begin{figure}
\noindent \begin{centering}
\includegraphics{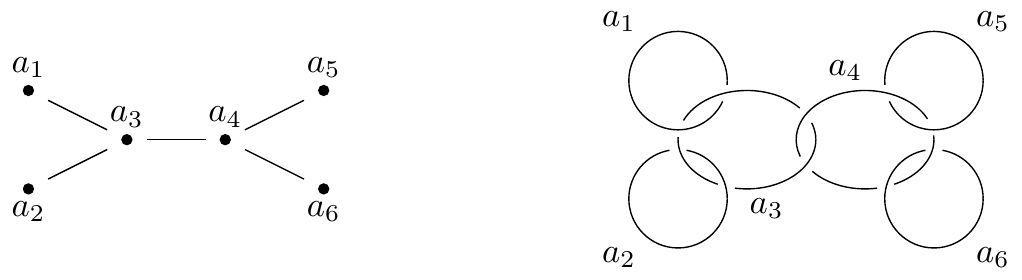}
\par\end{centering}
\caption{An example of a plumbing graph $\Gamma$ (left) and the associated
link of unknots (right) denoted as $\mathcal{L}(\Gamma)$. Each vertex
label corresponds to the framing of the respective link. The manifold
$M_{3}(\Gamma)$ can be constructed by performing a Dehn surgery on
$\mathcal{L}(\Gamma)$.\label{fig:Plumbing graph}}

\end{figure}

In this paper we focus on a very large class of 3-manifolds corresponding
to decorated graphs which, for simplicity, are assumed to be connected.
For a\,given graph $\Gamma$ we can obtain the associated plumbed
3-manifold $M_{3}(\Gamma)$ by performing a\,Dehn surgery on $\mathcal{L}(\Gamma)$
-- the corresponding link of framed unknots (see Figure~\ref{fig:Plumbing graph}).
We are mainly interested in Seifert fibrations over $S^{2}$ which
correspond to star-shaped graphs and are denoted by $M\left(b;\left\{ b_{i}/a_{i}\right\} _{i}\right)$,
where $b,b_{i},a_{i}\in\mathbb{Z}$. Among them there is a special
class of Brieskorn homology spheres. They are defined as the inetrsection
of the complex unit sphere with the\,hypersurface $z_{1}^{p_{1}}+z_{2}^{p_{2}}+z_{3}^{p_{3}}=0$
($p_{1},p_{2},p_{3}$ are coprime integers) and denoted by $\Sigma(p_{1},p_{2},p_{3})$.

Let us denote the set of vertices of $\Gamma$ by $V$ and the set
of edges by $E$. $L=|V|$ is equal to the\,number of components
of $\mathcal{L}(\Gamma)$. We can encode the information given by
the plumbing graph in a convenient way by the following $L\times L$
matrix

\begin{equation}
M_{v_{1},v_{2}}=\begin{cases}
1 & v_{1}\textrm{ and }v_{2}\textrm{ connected by the edge,}\\
a_{v} & v_{1}=v_{2}=v\textrm{ (framing of the link }v),\\
0 & \textrm{else}.
\end{cases}\qquad v_{i}\in V\cong\{1,\ldots,L\}
\end{equation}
From the link perspective $M$ is the linking matrix of $\mathcal{L}(\Gamma)$.
The cokernel of $M$ is equal (setwise) to the first homology group
of $M_{3}(\Gamma)$
\begin{equation}
H_{1}\left(M_{3}(\Gamma),\mathbb{Z}\right)\cong\textrm{Coker}M=\mathbb{Z}^{L}/M\mathbb{Z}^{L}.
\end{equation}
The number of elements in each set is given by $\det M$.

\subsection{Formula for RT invariants \label{subsec:RT-invariants-from-ST}}

In Appendix A of \cite{GPPV1701} the reasoning leading to equation
(\ref{eq:WRT from Zhat intro}) starts from the following formula
for the\,WRT invariant of a\,plumbed 3-manifold $M_{3}(\Gamma)$
\begin{align}
\textrm{WRT}[M_{3}(\Gamma);1/k]= & \frac{F[\Gamma;1/k]}{F[+1\bullet;1/k]^{b_{+}}F[-1\bullet;1/k]^{b_{-}}},\nonumber \\
F[\Gamma;1/k]= & \sum_{n\in\{1,\ldots,k-1\}^{L}}\prod_{v\in\textrm{V}}q^{\frac{a_{v}(n_{v}^{2}-1)}{4}}\left(q^{\frac{n_{v}}{2}}-q^{-\frac{n_{v}}{2}}\right)^{2-\deg(v)}\label{eq:WRT from ST}\\
 & \times\left.\frac{\prod_{(v_{1},v_{2})\in E}\left(q^{\frac{n_{v_{1}}n_{v_{2}}}{2}}-q^{-\frac{n_{v_{1}}n_{v_{2}}}{2}}\right)}{\left(q^{1/2}-q^{-1/2}\right)^{L+1}}\right|_{q=e^{\frac{2\pi i}{k}}},\nonumber 
\end{align}
where $b_{+}$ and $b_{-}$ are the number of positive and negative
eigenvalues of the matrix $M$. The symbol $\pm1\bullet$ denotes
the plumbing graph with one vertex corresponding to the unknot with
$\pm1$ framing. In this paper we always assume
\begin{equation}
q=e^{2\pi i\tau}\label{eq:q - tau}
\end{equation}
and the WRT invariant corresponds to $\tau=1/k$.

Equation (\ref{eq:WRT from ST}) comes from the quantum group construnction
\cite{RT91} where all roots of unity are on the same footing. More
formally, formula (\ref{eq:WRT from ST}) transforms equivariantly
under the Galois group $\textrm{Gal}\left(\mathbb{Q}(e^{2\pi i\frac{r}{s}})/\mathbb{Q}\right)$
\cite{LR99,LZ99} and in consequence its generalisation to $\tau=r/s$
is given by substitution $q=e^{2\pi i\frac{r}{s}}$

\begin{align}
\textrm{RT}[M_{3}(\Gamma);r/s]= & \frac{F[\Gamma;r/s]}{F[+1\bullet;r/s]^{b_{+}}F[-1\bullet;r/s]^{b_{-}}},\nonumber \\
F[\Gamma;r/s]= & \sum_{n\in\{1,\ldots,s-1\}^{L}}\prod_{v\in\textrm{V}}q^{\frac{a_{v}(n_{v}^{2}-1)}{4}}\left(q^{\frac{n_{v}}{2}}-q^{-\frac{n_{v}}{2}}\right)^{2-\deg(v)}\label{eq:RT from ST}\\
 & \times\left.\frac{\prod_{(v_{1},v_{2})\in E}\left(q^{\frac{n_{v_{1}}n_{v_{2}}}{2}}-q^{-\frac{n_{v_{1}}n_{v_{2}}}{2}}\right)}{\left(q^{1/2}-q^{-1/2}\right)^{L+1}}\right|_{q=e^{2\pi i\frac{r}{s}}}.\nonumber 
\end{align}
We will use this formula in many examples in Section \ref{sec:Examples},
but it is interesting on its own.

According to Turaev construction \cite{Tur94} we can associate a
modular tensor category (MTC) to the\,3d\,topological quantum field
theory. The MTC comes equipped with modular $S$ and $T$ matrices
which capture the structure of the topological partition function.
For the plumbed 3-manifold this relation reads (see \cite{Jeff92,DGNP1809}
for more details)
\begin{equation}
Z_{\textrm{top}}[M_{3}(\Gamma)]=\sum_{n}\prod_{v\in\textrm{V}}\left(T_{n_{v}n_{v}}\right)^{a_{v}}\left(S_{0n_{v}}\right)^{2-\deg(v)}\prod_{(v_{1},v_{2})\in E}S_{n_{v_{1}}n_{v_{2}}}.\label{eq:Ztop from S =000026 T matrices}
\end{equation}
Comparing (\ref{eq:RT from ST}) with (\ref{eq:Ztop from S =000026 T matrices})
we can see that the expression for $F$ matches the structure of $Z_{\textrm{top}}$
for
\begin{equation}
\begin{split}T_{mn} & =\delta_{m,n}q^{\frac{n^{2}-1}{4}},\\
S_{0n} & =\frac{1}{i\sqrt{2s}}\left(q^{\frac{n}{2}}-q^{-\frac{n}{2}}\right),\\
S_{mn} & =\frac{1}{i\sqrt{2s}}\left(q^{\frac{nm}{2}}-q^{-\frac{nm}{2}}\right),\\
q & =e^{2\pi i\frac{r}{s}}.
\end{split}
\label{eq:F from ST}
\end{equation}
This is a projective representation of $\textrm{SL}(2,\mathbb{Z})$,
where the phase factor is an integer multiple of $1/8$. In order
to restore $(ST)^{3}=\pm1$ we have to rescale $T$
\begin{equation}
T_{mn}\mapsto\delta_{m,n}q^{\frac{2n^{2}-s}{8}}.
\end{equation}
The condition $S^{2}=\pm1$ is ensured by the normalisation factor
$\frac{1}{i\sqrt{2s}}$ which cancels out in (\ref{eq:RT from ST}).

Another important observation is the invariance of formula (\ref{eq:RT from ST})
under $r\mapsto r+ns$ symmetry ($n\in\mathbb{Z}$). It is equivalent
to the multiplication of every $q$ by $e^{2\pi in}=1$. The $r\mapsto r+ns$
symmetry helps to solve the problem of choosing the branch of the
complex root which arises in the context of RT invariants (see Section
\ref{subsec:RT-invariants-from-Zhat}).

\section{Main conjecture \label{sec:Main-conjecture}}

\subsection{RT invariants from $\hat{Z}$ invariants \label{subsec:RT-invariants-from-Zhat}}

The reasoning leading to our main conjecture follows the Appendix
A of \cite{GPPV1701}, which starts from expression (\ref{eq:WRT from ST})
and, in the\,crucial step, uses the Gauss sum reciprocity formula
\begin{align}
\sum_{n\in\mathbb{Z}^{L}/2k\mathbb{Z}^{L}}\exp\left[\frac{\pi i}{2k}(n,Mn)+\frac{\pi i}{k}(l,n)\right] & =\label{eq:Gauss reciprocity}\\
\frac{e^{\pi i\sigma}(2k)^{L/2}}{|\det M|^{1/2}}\sum_{a\in\mathbb{Z}^{L}/M\mathbb{Z}^{L}} & \exp\left[-2\pi ik\left(a+\frac{l}{2k},M^{-1}\left[a+\frac{l}{2k}\right]\right)\right],\nonumber 
\end{align}
where $l\in\mathbb{Z}^{L}$, $(\cdot,\cdot)$ is the standard pairing
on $\mathbb{Z}^{L}$ and $\sigma=b_{+}-b_{-}$ is the signature of
the linking matrix $M$. The final result is the relation between
the WRT invariant and the $\hat{Z}$ invariant for $\tau=1/k$
\begin{align}
\textrm{WRT}[M_{3}(\Gamma);1/k]= & \underset{q\rightarrow e^{\frac{2\pi i}{k}}}{\lim}\frac{\sum_{a\in\textrm{Coker}M}e^{-2\pi ik(a,M^{-1}a)}\sum_{b\in2\textrm{Coker}M+\delta}S_{ab}\hat{Z}_{b}}{2\left(q^{1/2}-q^{-1/2}\right)},\nonumber \\
S_{ab}= & \frac{e^{-2\pi i(a,M^{-1}b)}}{|\det M|^{1/2}},\label{eq:WRT from Zhat}
\end{align}
where $\delta\in\mathbb{Z}^{L}/2\mathbb{Z}^{L}$ and $\delta_{v}\equiv\deg v\;\textrm{mod}\ 2$.

We would like to have an analogous derivation for $\tau=r/s$, so
we start from equation (\ref{eq:RT from ST}) and follow all the steps
of the Appendix A. The crucial one is again the Gauss sum reciprocity
formula. In order to deal with $\tau=r/s$ we have to rescale the
formula, which is equivalent to considering (\ref{eq:Gauss reciprocity})
for $\tilde{M}=rM$ and $\tilde{l}=rl$ (we also write $s$ instead
of $k$). We obtain
\begin{align}
\sum_{n\in\mathbb{Z}^{L}/2s\mathbb{Z}^{L}}\exp\left[\frac{\pi i}{2s}(n,rMn)+\frac{\pi i}{s}(rl,n)\right] & =\label{eq:Rescaled Gauss reciprocity}\\
\frac{e^{\pi i\sigma}(2s/r)^{L/2}}{|\det M|^{1/2}}\sum_{a\in\mathbb{Z}^{L}/rM\mathbb{Z}^{L}} & \exp\left[-2\pi is\left(a+\frac{rl}{2s},(rM)^{-1}\left[a+\frac{rl}{2s}\right]\right)\right],\nonumber 
\end{align}
which leads to our main conjecture
\begin{align}
\textrm{RT}[M_{3}(\Gamma);r/s]= & \underset{q\rightarrow e^{2\pi i\frac{r}{s}}}{\lim}\frac{\sum_{a\in\textrm{Coker}(rM)}e^{-2\pi i\frac{s}{r}(a,M^{-1}a)}\sum_{b\in2\textrm{Coker}M+\delta}S_{ab}\hat{Z}_{b}}{2\left(q^{1/2}-q^{-1/2}\right)G(s,r)^{L}},\nonumber \\
S_{ab}= & \frac{e^{-2\pi i(a,M^{-1}b)}}{|\det M|^{1/2}},\label{eq:Main conjecture - RT from Zhat}\\
G(s,r)= & \sum_{c\in\mathbb{Z}_{r}}e^{2\pi i\frac{s}{r}c^{2}}=\begin{cases}
\sqrt{r}\left(\frac{s}{r}\right) & r\equiv1\;\textrm{mod}\ 4\\
i\sqrt{r}\left(\frac{s}{r}\right) & r\equiv3\;\textrm{mod}\ 4
\end{cases}\nonumber 
\end{align}
where $\left(\frac{s}{r}\right)$ is the Jacobi symbol. If we want
to use above formula for even $r$, we have to choose another representant
of the $r\sim r+ns$ equivalence class to avoid dividing by $G(s,r)=0$
(in fact this happens only for $r\equiv2\;\textrm{mod}\ 4$ but it
is more convenient to treat all even $r$ the same). This problem
is a reflection of the fact that for some choices of roots of $q=e^{2\pi i\frac{r}{s}}$
(for $SU(2)$ we deal with 4 values of\,$q^{1/4}$) we have $F[\pm1\bullet;r/s]=0$.
A detailed discussion of the vanishing denominator in the RT invariants
can be found in \cite{Le03,HL1503}.

There are two important differences between (\ref{eq:Main conjecture - RT from Zhat})
and (\ref{eq:WRT from Zhat}). The first one is in the summation range
-- $\textrm{Coker}(rM)$ has $r^{L}$ more elements than $\textrm{Coker}M$.
On the other hand we have $G(s,r)^{L}$ in denominator which scales
as $r^{L/2}$ and ``compensates'' this growth. For $r=1$ equation
(\ref{eq:Main conjecture - RT from Zhat}) reduces to (\ref{eq:WRT from Zhat})
which provides the first consistency check.

\subsection{Rational $\tau$ limit of $\hat{Z}$ invariants\label{subsec:Rational--limit of Zhat}}

For some simple 3-manifolds such as lens spaces $L(p,1)$ the $\tau\rightarrow r/s$
limit of the $\hat{Z}$ invariant is very easy to obtain (see Section
\ref{subsec:Lens-spaces}), however these are exceptions rather than
the rule. Fortunately for many 3-manifolds (e.g. Seifert manifolds
with 3 singular fibers) the $\hat{Z}$ invariant can be expressed
as a\,linear combination of false theta functions defined as
\begin{equation}
\begin{split}\Psi_{m,\alpha} & =\sum_{n=0}^{\infty}\psi_{2m,\alpha}(n)q^{\frac{n^{2}}{4m}}=\sum_{n=0}^{\infty}\psi_{2m,\alpha}(n)e^{\frac{\pi i\tau n^{2}}{2m}},\\
\psi_{2m,\alpha}(n) & =\begin{cases}
\pm1 & n\equiv\pm\alpha\;\textrm{mod}\ 2m,\\
0 & \textrm{otherwise}.
\end{cases}
\end{split}
\label{eq:False theta def}
\end{equation}
In this case the calculation of $\underset{\tau\rightarrow r/s}{\lim}\hat{Z}$
is more difficult, but still possible. In \cite{LZ99,Hik0305} we
find that
\begin{equation}
\underset{\tau\rightarrow r/s}{\lim}\Psi_{m,\alpha}=\sum_{n=0}^{ms}\psi_{2m,\alpha}(n)\left(1-\frac{1}{ms}\right)e^{\frac{\pi irn^{2}}{2ms}}.\label{eq:Hikami's limit}
\end{equation}
Since this result is an essential tool in Section \ref{sec:Examples},
it serves also as the guiding rule in choosing examples for testing
our main conjecture.

\subsection{Conventions \label{subsec:Conventions}}

Before moving to examples let us discuss some conventional issues.

In many papers, e.g. \cite{GPV1602,GMP1605,CCFGH1809}, the normalisation
of the RT invariant (or the WRT invariant for\,$\tau=1/k$) is different.
In our notation
\begin{equation}
\textrm{RT}[S^{3};r/s]=1,
\end{equation}
whereas there
\begin{equation}
\textrm{RT}_{\textrm{CS}}[S^{2}\times S^{1};r/s]=1.
\end{equation}
We write $\textrm{RT}_{\textrm{CS}}$ because this notation is based
on the value of the Chern-Simons partition function for $r/s=1/k$
(many authors write $Z_{\textrm{CS}}$ instead of $\textrm{RT}_{\textrm{CS}}$
but we want to avoid the confusion with $\hat{Z}$). The relation
between these two conventions is given by
\begin{equation}
\textrm{RT}[M_{3}(\Gamma);r/s]=\frac{i\sqrt{2s}}{q^{1/2}-q^{-1/2}}\textrm{RT}_{\textrm{CS}}[M_{3}(\Gamma);r/s].
\end{equation}

The second issue is related to the $\mathbb{Z}_{2}$ symmetry group
acting on $\textrm{Coker}M\cong H_{1}\left(M_{3}(\Gamma),\mathbb{Z}\right)$
by $a\mapsto-a$. Since (\ref{eq:Main conjecture - RT from Zhat})
is invariant under this transformation and $\hat{Z}_{a}=\hat{Z}_{-a}$
we could write
\begin{align}
\textrm{RT}[M_{3}(\Gamma);r/s]= & \underset{q\rightarrow e^{2\pi i\frac{r}{s}}}{\lim}\frac{\sum_{a\in\textrm{Coker}(rM)/\mathbb{Z}_{2}}e^{-2\pi i\frac{s}{r}(a,M^{-1}a)}\sum_{b\in\left(2\textrm{Coker}M+\delta\right)/\mathbb{Z}_{2}}S'_{ab}\hat{Z}'_{b}}{2\left(q^{1/2}-q^{-1/2}\right)G(s,r)^{L}},\nonumber \\
S'_{ab}= & \frac{\sum_{a'\in\left\{ \mathbb{Z}_{2}\textrm{-orbit of }a\right\} }e^{-2\pi i(a',M^{-1}b)}}{|\det M|^{1/2}},\label{eq:RT from Zhat unfolded}\\
\hat{Z}'_{b}= & \left|\mathbb{Z}_{2}\textrm{-orbit of }b\right|\hat{Z}{}_{b}.\nonumber 
\end{align}
This convention is often called \emph{folded} whereas ours -- \emph{unfolded}.
The former is present in \cite{GPV1602,GMP1605,GPPV1701,CCFGH1809},
we use the latter because it is inconvenient to divide $\textrm{Coker}(rM)$
by $\mathbb{Z}_{2}$ for every considered $r$. We would like to stress
that because of that our $\hat{Z}_{b}$ differs from the folded one
(denoted by $\hat{Z}'_{b}$) by the factor of $2$ if $b$ is not
a fixed point of $\mathbb{Z}_{2}$ symmetry. Moreover, some papers
use different numeration of $\hat{Z}_{b}$. Detailed discussion of
this issue can be found in \cite{GPPV1701}.

\section{Examples \label{sec:Examples}}

In this section we test our main conjecture (\ref{eq:Main conjecture - RT from Zhat})
by comparing it to (\ref{eq:RT from ST}) on various examples. All
computations are done numerically using Mathematica.

\subsection{Lens spaces $L(p,1)$\label{subsec:Lens-spaces}}

\noindent For the lens space $L(p,1)$ the plumbing graph $\Gamma$
is given by 
\begin{figure}[H]
\noindent \centering{}\includegraphics[height=0.5cm]{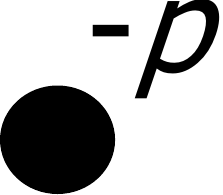}.
\end{figure}
 In consequence $L=1$, $M=\left[-p\right]$, $\textrm{Coker}(rM)=\mathbb{Z}_{rp}$,
and $\textrm{2Coker}M+\delta=2\mathbb{Z}_{p}$. However, only for
three $b\in2\mathbb{Z}_{p}$ the invariant $\hat{Z}_{b}$ is non-zero
\cite{GPV1602}
\begin{equation}
\hat{Z}_{0}=-2q^{\frac{p-3}{4}},\qquad\hat{Z}_{-2}=\hat{Z}_{2}=q^{\frac{p-3}{4}}q^{\frac{1}{p}}.
\end{equation}
Therefore the formula (\ref{eq:Main conjecture - RT from Zhat}) reduces
to
\begin{equation}
\textrm{RT}[L(p,1);r/s]=\underset{q\rightarrow e^{2\pi i\frac{r}{s}}}{\lim}\frac{\sum_{a\in\mathbb{Z}_{rp}}e^{2\pi i\frac{s}{r}\frac{a^{2}}{p}}\sum_{b\in\{-2,0,2\}}e^{2\pi i\frac{ab}{p}}\hat{Z}_{b}}{2\sqrt{p}\left(q^{1/2}-q^{-1/2}\right)G(s,r)}.\label{eq:RT for Lens space general}
\end{equation}
On the other hand we can use (\ref{eq:RT from ST}) to write
\begin{equation}
\textrm{RT}[L(p,1);r/s]=\left.\frac{\sum_{n\in\{1,\ldots,s-1\}}q^{(-p)\frac{n^{2}-1}{4}}\left(q^{\frac{n}{2}}-q^{-\frac{n}{2}}\right)^{2}}{\sum_{n\in\{1,\ldots,s-1\}}q^{(-1)\frac{n^{2}-1}{4}}\left(q^{\frac{n}{2}}-q^{-\frac{n}{2}}\right)^{2}}\right|_{q=e^{2\pi i\frac{r}{s}}}.\label{eq:RT for Lens space from ST}
\end{equation}
Using Mathematica we checked that (\ref{eq:RT for Lens space general})
and (\ref{eq:RT for Lens space from ST}) give the same result. We
compared both formulas for $p=3,5,7,9,11$ and $r/s$ up to $16/17$.

\subsection{Brieskorn spheres\label{subsec:Brieskorn-spheres}}

Brieskorn homology spheres $\Sigma(p_{1},p_{2},p_{3})$ are interesting
examples, because in their case $\textrm{Coker}M=\{0\}$ so we have
only one invariant $\hat{Z}_{b}=\hat{Z}_{\delta}$ and the RT invariant
is equal (up to normalisation) to $\hat{Z}_{\delta}$ \cite{CCFGH1809}

\begin{equation}
\textrm{RT}[\Sigma(p_{1},p_{2},p_{3});r/s]=\underset{q\rightarrow e^{2\pi i\frac{r}{s}}}{\lim}\frac{\hat{Z}_{\delta}}{2\left(q^{1/2}-q^{-1/2}\right)}.\label{eq:RT from Zhat Brieskorn}
\end{equation}
For $r=1$ this statement immediately follows from (\ref{eq:WRT from Zhat}).
However $\hat{Z}_{\delta}/2\left(q^{1/2}-q^{-1/2}\right)$ is defined
for all $\tau\in\mathbb{H}$ ($q$ inside unit disk) with well-defined
limits at all rational $\tau$, so in this case there is no difference
between $r=1$ and other integers. Comparing (\ref{eq:RT from Zhat Brieskorn})
with (\ref{eq:Main conjecture - RT from Zhat}) we can see that

\begin{equation}
\frac{\sum_{a\in\textrm{Coker}(rM)}e^{-2\pi i\frac{s}{r}(a,M^{-1}a)}S_{a\delta}}{G(s,r)^{L}}=1,\label{eq:Brieskorn simplification}
\end{equation}
which we numerically checked using Mathematica.

\subsubsection{$\Sigma(2,3,7)$}

The graph $\Gamma_{\Sigma(2,3,7)}$ of the $\Sigma(2,3,7)$ Brieskorn
sphere is given by 
\begin{figure}[H]
\noindent \centering{}\includegraphics[height=2cm]{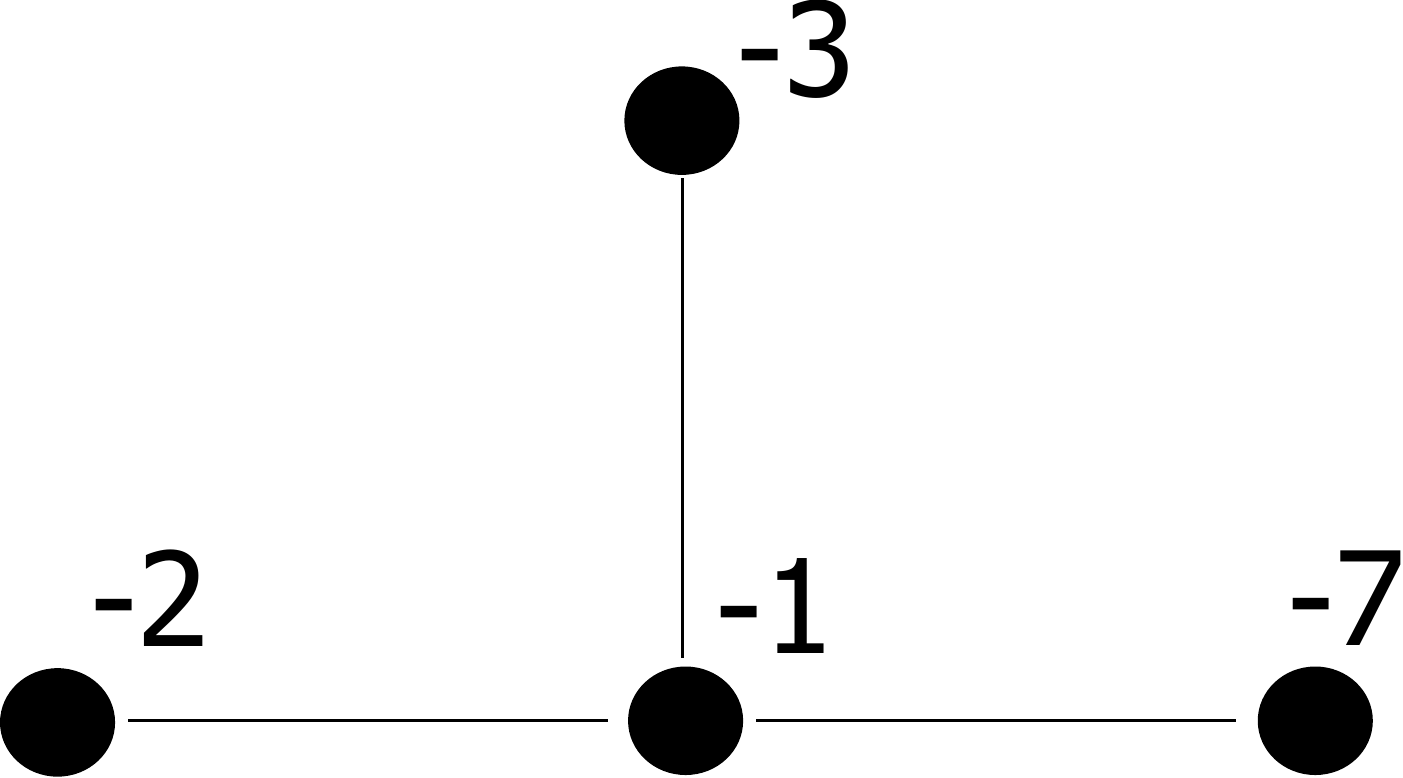}.
\end{figure}
 We number vertices in the following way (we do it for all 4-vertex
graphs in this paper) 
\begin{figure}[H]
\noindent \centering{}\includegraphics[height=2cm]{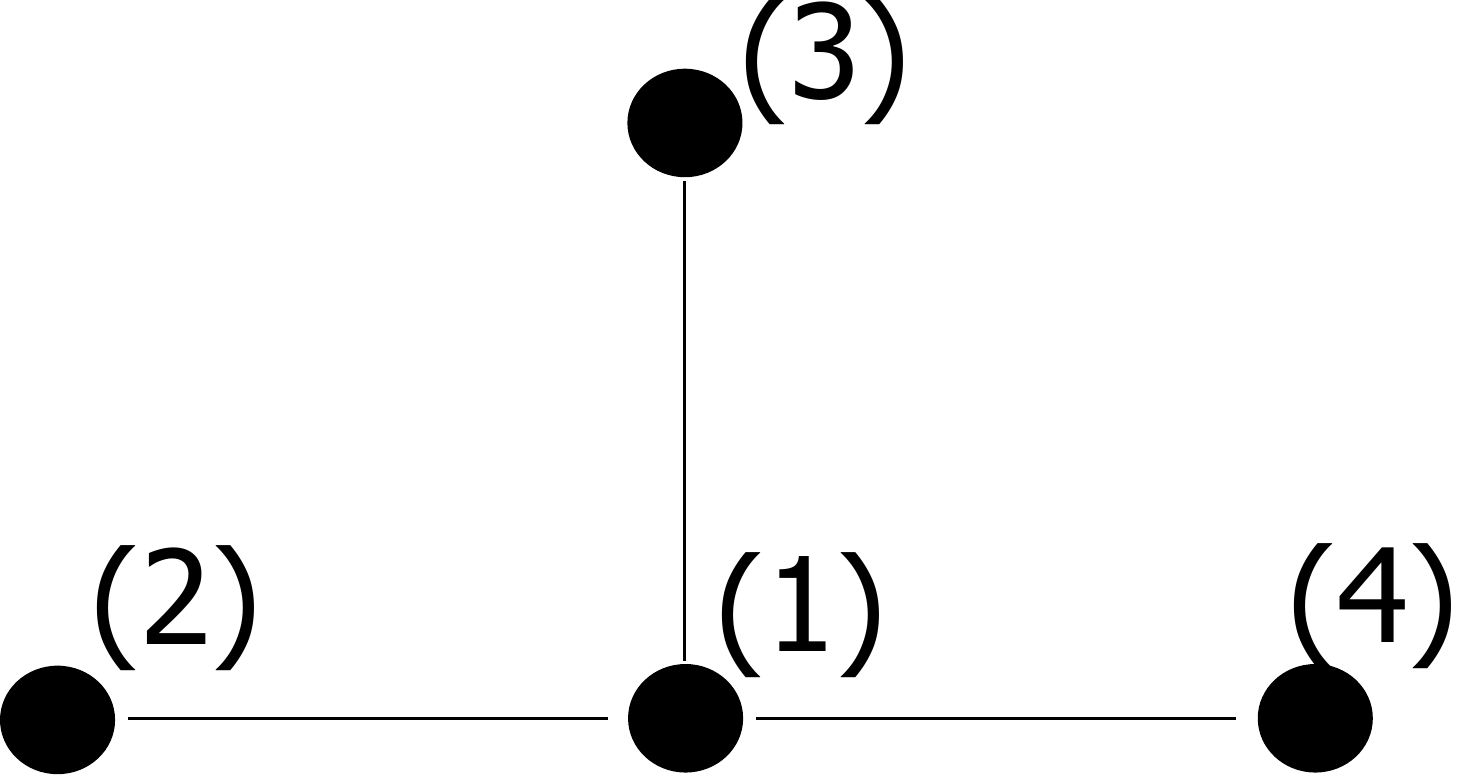}.
\end{figure}
 In consequence the linking matrix reads
\begin{equation}
M=\left[\begin{array}{cccc}
-1 & 1 & 1 & 1\\
1 & -2 & 0 & 0\\
1 & 0 & -3 & 0\\
1 & 0 & 0 & -7
\end{array}\right],
\end{equation}
so $\det M=1$ and $\textrm{Coker}M=\{0\}$. $\hat{Z}_{\delta}$ is
given by \cite{GMP1605}
\begin{equation}
\hat{Z}_{\delta}=q^{\frac{83}{168}}\left(\Psi_{42,1}-\Psi_{42,13}-\Psi_{42,29}+\Psi_{42,41}\right)\label{eq:Zhatdelta Sigma(2,3,7)}
\end{equation}
 (There is a typo in \cite{GMP1605}, $q^{\frac{83}{168}}$ should
be in numerator as in (\ref{eq:Zhatdelta Sigma(2,3,7)})). Therefore
\begin{equation}
\textrm{RT}[\Sigma(2,3,7);r/s]=\frac{\left.\hat{Z}_{\delta}\right|_{\tau=r/s}}{4i\sin\left(\pi\frac{r}{s}\right)},\label{eq:RT for Sigma(2,3,7) general}
\end{equation}
where
\begin{equation}
\left.\hat{Z}_{\delta}\right|_{\tau=r/s}=e^{\frac{83}{84}\pi i\frac{r}{s}}\sum_{n=0}^{42s}\left(\psi_{84,1}(n)-\psi_{84,13}(n)-\psi_{84,29}(n)+\psi_{84,41}(n)\right)\left(1-\frac{1}{42s}\right)e^{\frac{\pi ir}{84s}n^{2}}
\end{equation}
was calculated by applying (\ref{eq:Hikami's limit}) to (\ref{eq:Zhatdelta Sigma(2,3,7)}).

The formula (\ref{eq:RT from ST}) gives
\begin{align}
\textrm{RT}[\Sigma(2,3,7);r/s]= & \frac{F[\Gamma_{\Sigma(2,3,7)};r/s]}{F[-1\bullet;r/s]^{4}},\nonumber \\
F[-1\bullet;r/s]= & \left.\frac{\sum_{n\in\{1,\ldots,s-1\}}q^{(-1)\frac{n^{2}-1}{4}}\left(q^{\frac{n}{2}}-q^{-\frac{n}{2}}\right)^{2}}{\left(q^{1/2}-q^{-1/2}\right)^{2}}\right|_{q=e^{2\pi i\frac{r}{s}}},\label{eq:RT for Sigma(2,3,7) from ST}\\
F[\Gamma_{\Sigma(2,3,7)};r/s]= & \left.\frac{\sum_{n\in\{1,\ldots,s-1\}^{4}}\prod_{v\in\textrm{V}}T_{n_{v}n_{v}}^{a_{v}}S_{0n_{v}}^{2-\deg(v)}\prod_{(v_{1},v_{2})\in E}S_{n_{v_{1}}n_{v_{2}}}}{\left(q^{1/2}-q^{-1/2}\right)^{5}}\right|_{q=e^{2\pi i\frac{r}{s}}},\nonumber 
\end{align}
where\footnote{For simplicity we do not include the $\frac{1}{i\sqrt{2s}}$ prefactor
in formulas for $S$ matrices in the whole Section \ref{sec:Examples}.}
\begin{align}
T_{n_{v}n_{v}}^{a_{v}}= & q^{\frac{a_{v}(n_{v}^{2}-1)}{4}},\quad a_{v}=\begin{cases}
-1 & v=1\\
-2 & v=2\\
-3 & v=3\\
-7 & v=4
\end{cases}\nonumber \\
S_{n_{v_{1}}n_{v_{2}}}= & q^{\frac{n_{v_{1}}n_{v_{2}}}{2}}-q^{-\frac{n_{v_{1}}n_{v_{2}}}{2}},\quad(v_{1},v_{2})=(1,2),(1,3),(1,4)\label{eq:S =000026 T matrices Sigma(2,3,7)}\\
S_{0n_{v}}^{2-\deg(v)}= & \begin{cases}
\left[q^{\frac{n_{v}}{2}}-q^{-\frac{n_{v}}{2}}\right]^{-1} & v=1\\
1 & v=2,3,4
\end{cases}\nonumber 
\end{align}
Using Mathematica we checked -- for all $r/s$ up to $12/13$ --
that (\ref{eq:RT for Sigma(2,3,7) general}) and (\ref{eq:RT for Sigma(2,3,7) from ST})
give the same result.

\subsubsection{Poncar\'{e} sphere}

For the Poincar\'{e} sphere $\Sigma(2,3,5)$ we have the following
plumbing graph $\Gamma_{\Sigma(2,3,5)}$ 
\begin{figure}[H]
\noindent \centering{}\includegraphics[height=2cm]{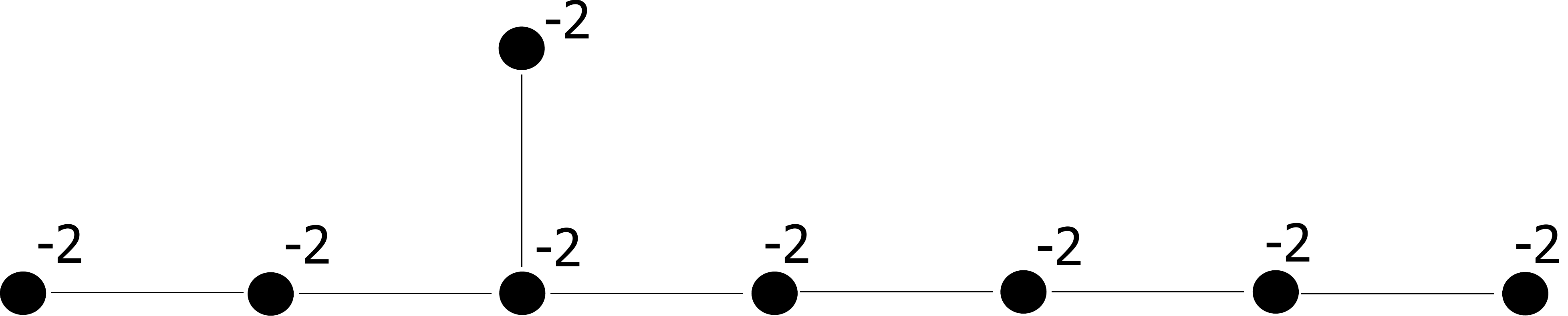}.
\end{figure}
 The numbering 
\begin{figure}[H]
\noindent \centering{}\includegraphics[height=2cm]{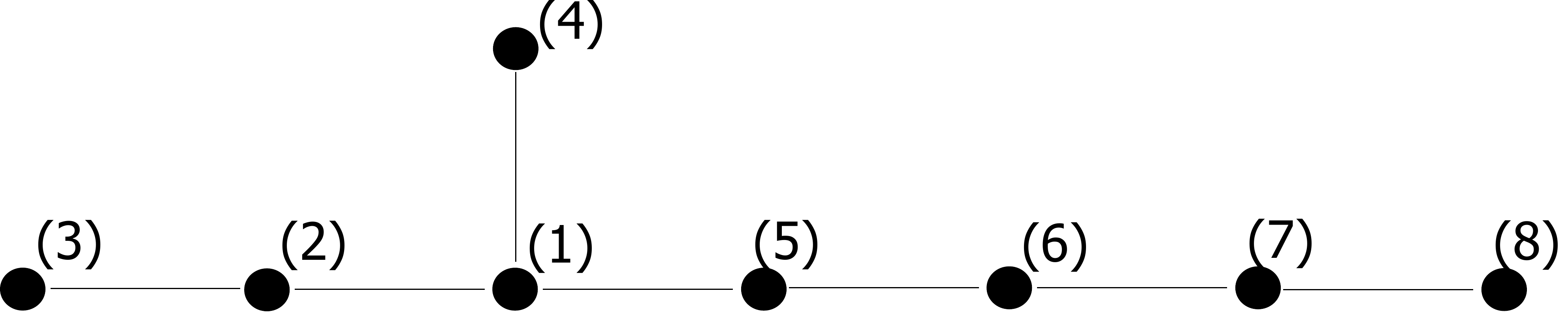}
\end{figure}
 leads to
\begin{equation}
M=\left[\begin{array}{cccccccc}
-2 & 1 & 0 & 1 & 1 & 0 & 0 & 0\\
1 & -2 & 1 & 0 & 0 & 0 & 0 & 0\\
0 & 1 & -2 & 0 & 0 & 0 & 0 & 0\\
1 & 0 & 0 & -2 & 0 & 0 & 0 & 0\\
1 & 0 & 0 & 0 & -2 & 1 & 0 & 0\\
0 & 0 & 0 & 0 & 1 & -2 & 1 & 0\\
0 & 0 & 0 & 0 & 0 & 1 & -2 & 1\\
0 & 0 & 0 & 0 & 0 & 0 & 1 & -2
\end{array}\right].
\end{equation}
We have $\det M=1$, $\textrm{Coker}M=\{0\}$ again and $\hat{Z}_{\delta}$
is given by \cite{GMP1605}
\begin{equation}
\hat{Z}_{\delta}=q^{-\frac{181}{120}}\left[2q^{\frac{1}{120}}-(\Psi_{30,1}+\Psi_{30,11}+\Psi_{30,19}+\Psi_{30,29})\right].\label{eq:Zhatdelta Poincare}
\end{equation}
Therefore
\begin{align}
\textrm{RT}[\Sigma(2,3,5);r/s]= & \frac{\left.\hat{Z}_{\delta}\right|_{\tau=r/s}}{4i\sin\left(\pi\frac{r}{s}\right)},\label{eq:RT for Poincare general}
\end{align}
where
\begin{equation}
\left.\hat{Z}_{\delta}\right|_{\tau=r/s}=e^{-\frac{181}{60}\pi i\frac{r}{s}}\left[2e^{\frac{\pi ir}{60s}}-\sum_{n=0}^{30s}\left(\psi_{60,1}(n)+\psi_{60,11}(n)+\psi_{60,19}(n)+\psi_{60,29}(n)\right)\left(1-\frac{1}{30s}\right)e^{\frac{\pi ir}{60s}n^{2}}\right].
\end{equation}

On the other hand equation (\ref{eq:RT from ST}) leads to
\begin{align}
\textrm{RT}[\Sigma(2,3,5);r/s]= & \frac{F[\Gamma_{\Sigma(2,3,5)};r/s]}{F[-1\bullet;r/s]^{8}},\label{eq:RT for Poincare from ST}\\
F[\Gamma_{\Sigma(2,3,5)};r/s]= & \left.\frac{\sum_{n\in\{1,\ldots,s-1\}^{8}}\prod_{v\in\textrm{V}}T_{n_{v}n_{v}}^{-2}S_{0n_{v}}^{2-\deg(v)}\prod_{(v_{1},v_{2})\in E}S_{n_{v_{1}}n_{v_{2}}}}{\left(q^{1/2}-q^{-1/2}\right)^{9}}\right|_{q=e^{2\pi i\frac{r}{s}}},\nonumber 
\end{align}
where
\begin{align}
T_{n_{v}n_{v}}^{-2}= & q^{\frac{1-n_{v}^{2}}{2}},\quad v=1,2,\ldots,8\nonumber \\
S_{n_{v_{1}}n_{v_{2}}}= & q^{\frac{n_{v_{1}}n_{v_{2}}}{2}}-q^{-\frac{n_{v_{1}}n_{v_{2}}}{2}},\;(v_{1},v_{2})=(1,2),(2,3),(1,4),(1,5),(5,6),(6,7),(7,8)\nonumber \\
S_{0n_{v}}^{2-\deg(v)}= & \begin{cases}
\left[q^{\frac{n_{v}}{2}}-q^{-\frac{n_{v}}{2}}\right]^{-1} & v=1\\
1 & v=2,5,6,7\\
q^{\frac{n_{v}}{2}}-q^{-\frac{n_{v}}{2}} & v=3,4,8.
\end{cases}
\end{align}
We have used Mathematica to check that (\ref{eq:RT for Poincare general})
and (\ref{eq:RT for Poincare from ST}) give the same result. Having
8 vertices was much more involved for the computer so we stopped at
$r/s=8/9$.

\subsection{Other Seifert manifolds\label{subsec:Other-Seifert-manifolds}}

\subsubsection{$M\left(-1;\frac{1}{2},\frac{1}{3},\frac{1}{9}\right)$ }

The Seifert manifold $M\left(-1;\frac{1}{2},\frac{1}{3},\frac{1}{9}\right)$
can be described by the plumbing graph $\Gamma_{M\left(-1;\frac{1}{2},\frac{1}{3},\frac{1}{9}\right)}$
\begin{figure}[H]
\noindent \centering{}\includegraphics[height=2cm]{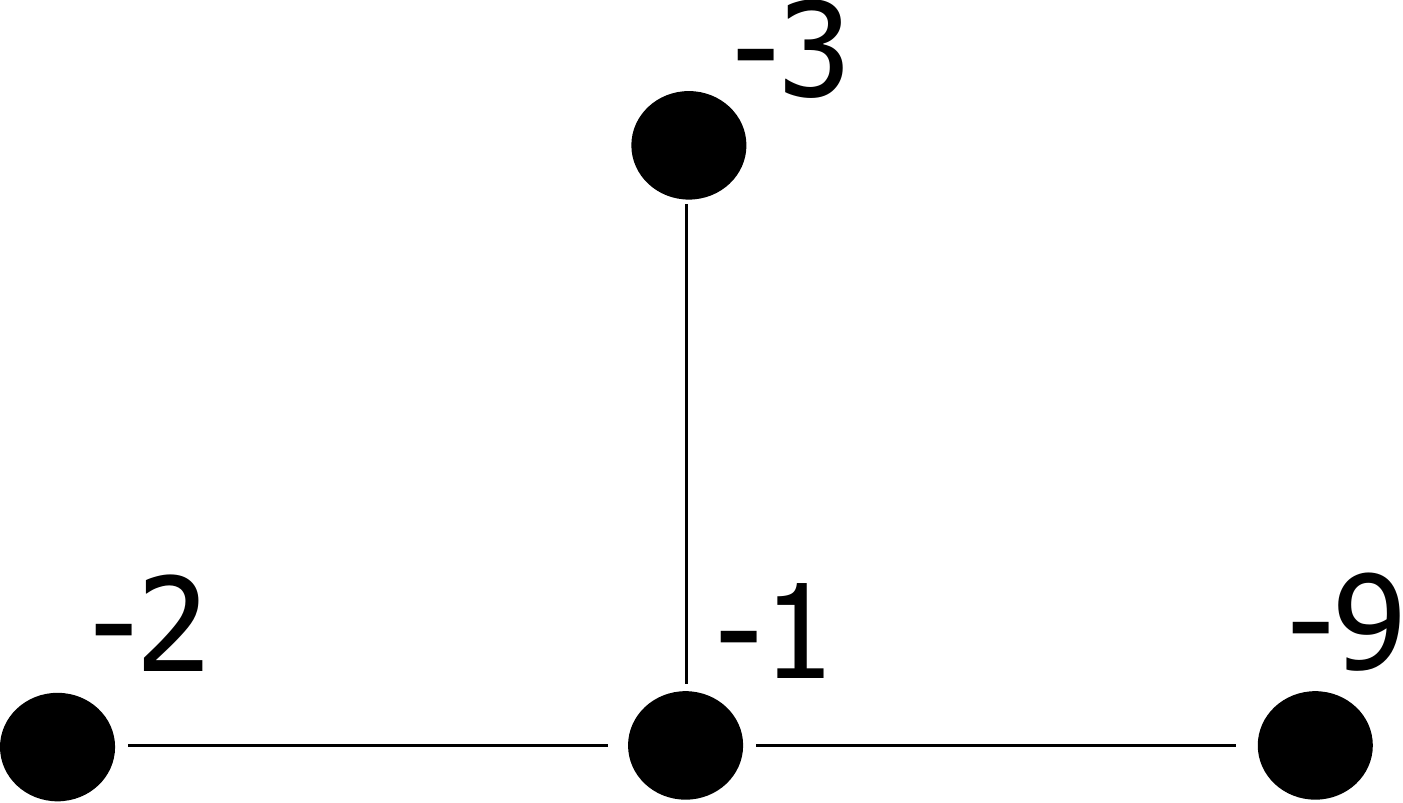}
\end{figure}
 and the linking matrix
\begin{equation}
M=\left[\begin{array}{cccc}
-1 & 1 & 1 & 1\\
1 & -2 & 0 & 0\\
1 & 0 & -3 & 0\\
1 & 0 & 0 & -9
\end{array}\right].
\end{equation}
Therefore $\det M=3$ and
\begin{equation}
\textrm{Coker}M=\left\{ \left[\begin{array}{c}
0\\
0\\
0\\
0
\end{array}\right],\left[\begin{array}{c}
1\\
0\\
-1\\
-6
\end{array}\right],\left[\begin{array}{c}
-1\\
0\\
1\\
6
\end{array}\right]\right\} .
\end{equation}
We have 
\begin{equation}
\delta=\left[\begin{array}{c}
1\\
-1\\
-1\\
-1
\end{array}\right]\Rightarrow b\in2\textrm{Coker}M+\delta=\left\{ \left[\begin{array}{c}
1\\
-1\\
-1\\
-1
\end{array}\right],\left[\begin{array}{c}
3\\
-1\\
-3\\
-13
\end{array}\right],\left[\begin{array}{c}
-3\\
1\\
3\\
13
\end{array}\right]\right\} 
\end{equation}
and $\hat{Z}$ invariants are given by \cite{CCFGH1809}
\begin{equation}
\begin{split}\hat{Z}_{\left[1,-1,-1,-1\right]}= & q^{71/72}\left(\Psi_{18,1}+\Psi_{18,17}\right),\\
\hat{Z}_{\left[-3,1,3,13\right]}=\hat{Z}_{\left[3,-1,-3,-13\right]}= & -\frac{1}{2}q^{71/72}\left(\Psi_{18,5}+\Psi_{18,13}\right).
\end{split}
\end{equation}
We can use (\ref{eq:Hikami's limit}) to compute $\left.\hat{Z}_{b}\right|_{\tau=r/s}$
and then (\ref{eq:Main conjecture - RT from Zhat}) leads to
\begin{align}
\textrm{RT}\left[M\left(-1;\frac{1}{2},\frac{1}{3},\frac{1}{9}\right);\frac{r}{s}\right]= & \frac{\sum_{a\in\textrm{Coker}(rM)}e^{-2\pi i\frac{s}{r}(a,M^{-1}a)}\sum_{b\in2\textrm{Coker}M+\delta}S_{ab}\left.\hat{Z}_{b}\right|_{\tau=r/s}}{4i\sin\left(\pi\frac{r}{s}\right)G(s,r)^{4}},\nonumber \\
S_{ab}= & \frac{e^{-2\pi i(a,M^{-1}b)}}{\sqrt{3}}.\label{eq:RT from Zhat for M(-1,1/2,1/3,1/9)}
\end{align}
In contrary to the Brieskorn spheres all terms are nontrivial.

On the other hand (\ref{eq:RT from ST}) gives
\begin{align}
\textrm{RT}\left[M\left(-1;\frac{1}{2},\frac{1}{3},\frac{1}{9}\right);\frac{r}{s}\right]= & \frac{F\left[\Gamma_{M\left(-1;\frac{1}{2},\frac{1}{3},\frac{1}{9}\right)};\frac{r}{s}\right]}{F[-1\bullet;\frac{r}{s}]^{4}},\label{eq:RT from ST for M(-1,1/2,1/3,1/9)}\\
F\left[\Gamma_{M\left(-1;\frac{1}{2},\frac{1}{3},\frac{1}{9}\right)};\frac{r}{s}\right]= & \left.\frac{\sum_{n\in\{1,\ldots,s-1\}^{4}}\prod_{v\in\textrm{V}}T_{n_{v}n_{v}}^{a_{v}}S_{0n_{v}}^{2-\deg(v)}\prod_{(v_{1},v_{2})\in E}S_{n_{v_{1}}n_{v_{2}}}}{\left(q^{1/2}-q^{-1/2}\right)^{5}}\right|_{q=e^{2\pi i\frac{r}{s}}},\nonumber 
\end{align}
where $S$ and $T$ matrices are the same as in (\ref{eq:S =000026 T matrices Sigma(2,3,7)})
except $a_{v}=-9$ for $v=4$.

We used Mathematica to check that (\ref{eq:RT from Zhat for M(-1,1/2,1/3,1/9)})
and (\ref{eq:RT from ST for M(-1,1/2,1/3,1/9)}) give the same result.
Because of the necessity of calculating $\textrm{Coker}(rM)$ for
each $r$ it was easier to increase the parameter $s$ and we stopped
at $r/s=7/30$.

\subsubsection{$M\left(-2;\frac{1}{2},\frac{1}{3},\frac{1}{2}\right)$}

The Seifert manifold $M\left(-2;\frac{1}{2},\frac{1}{3},\frac{1}{2}\right)$
has the following plumbing graph $\Gamma_{M\left(-2;\frac{1}{2},\frac{1}{3},\frac{1}{2}\right)}$
\begin{figure}[H]
\noindent \centering{}\includegraphics[height=2cm]{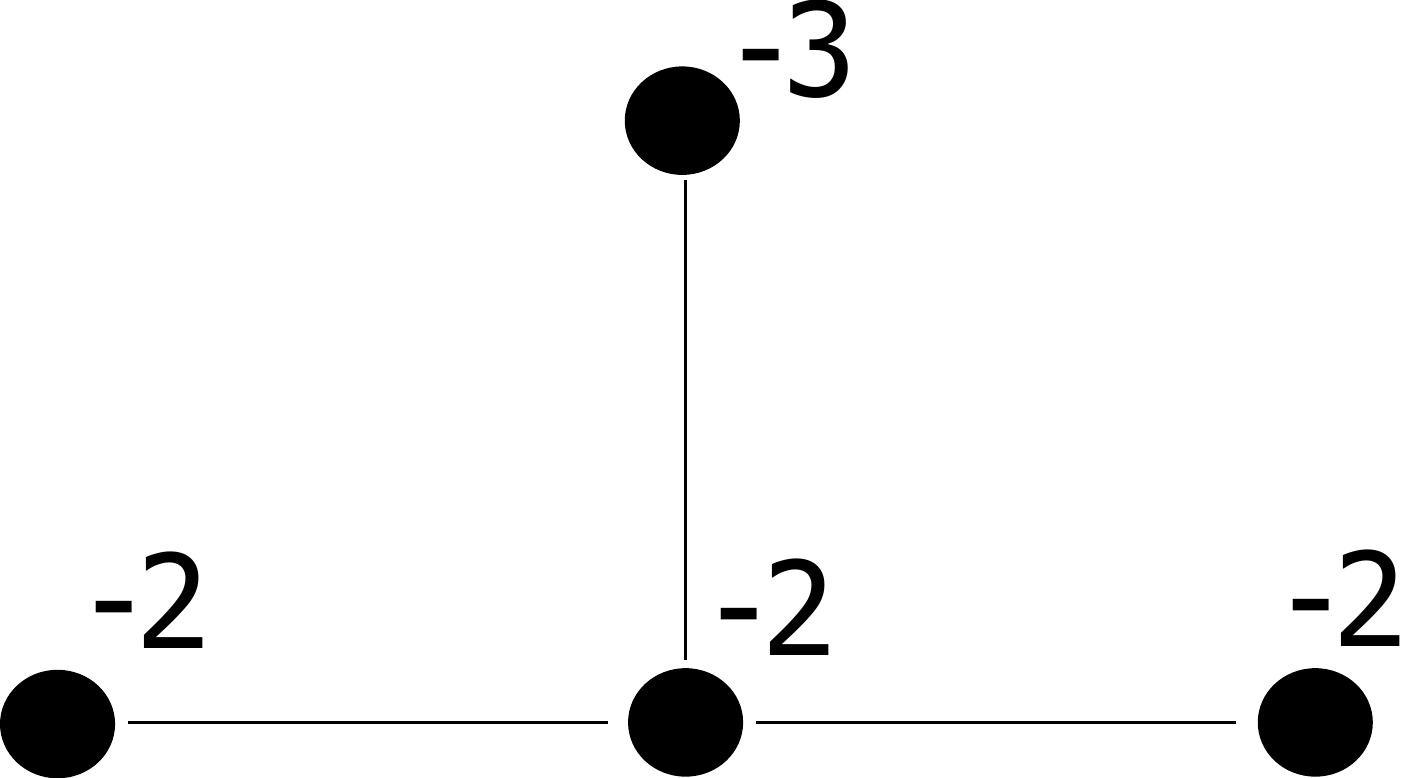}
\end{figure}
 and linking matrix
\begin{equation}
M=\left[\begin{array}{cccc}
-2 & 1 & 1 & 1\\
1 & -2 & 0 & 0\\
1 & 0 & -3 & 0\\
1 & 0 & 0 & -2
\end{array}\right].
\end{equation}
Therefore $\det M=8$, $\delta=\left[1,-1,-1,-1\right]$ and
\begin{equation}
b\in2\textrm{Coker}M+\delta=\left\{ \left[\begin{array}{c}
1\\
-1\\
-1\\
-1
\end{array}\right],\pm\left[\begin{array}{c}
3\\
-1\\
-5\\
-3
\end{array}\right],\pm\left[\begin{array}{c}
3\\
-3\\
-5\\
-1
\end{array}\right],\left[\begin{array}{c}
3\\
-3\\
-1\\
-3
\end{array}\right],\pm\left[\begin{array}{c}
1\\
-3\\
-1\\
-1
\end{array}\right]\right\} .
\end{equation}
$\hat{Z}$ invariants are given by \cite{CCFGH1809}
\begin{equation}
\begin{split}\hat{Z}_{\left[1,-1,-1,-1\right]}= & q^{-\frac{5}{12}}\left[2q^{\frac{1}{24}}-(\Psi_{6,1}+\Psi_{6,7})\right]\\
\hat{Z}_{\pm\left[3,-1,-5,-3\right]}=\hat{Z}_{\pm\left[3,-3,-5,-1\right]}= & -\frac{1}{2}q^{-\frac{5}{12}}\Psi_{6,2}\\
\hat{Z}_{\left[3,-3,-1,-3\right]}= & -q^{-\frac{5}{12}}(\Psi_{6,1}+\Psi_{6,7})\\
\hat{Z}_{\pm\left[1,-3,-1,-1\right]}= & q^{-\frac{5}{12}}\Psi_{6,4}
\end{split}
\end{equation}
Following the previous examples we use (\ref{eq:Hikami's limit})
to compute $\left.\hat{Z}_{b}\right|_{\tau=r/s}$ and then (\ref{eq:Main conjecture - RT from Zhat})
to obtain
\begin{align}
\textrm{RT}\left[M\left(-2;\frac{1}{2},\frac{1}{3},\frac{1}{2}\right);\frac{r}{s}\right]= & \frac{\sum_{a\in\textrm{Coker}(rM)}e^{-2\pi i\frac{s}{r}(a,M^{-1}a)}\sum_{b\in2\textrm{Coker}M+\delta}S_{ab}\left.\hat{Z}_{b}\right|_{\tau=r/s}}{4i\sin\left(\pi\frac{r}{s}\right)G(s,r)^{4}},\nonumber \\
S_{ab}= & \frac{e^{-2\pi i(a,M^{-1}b)}}{\sqrt{8}}.\label{eq:RT from Zhat for M(-2,1/2,1/3,1/2)}
\end{align}
Similarly to $M\left(-1;\frac{1}{2},\frac{1}{3},\frac{1}{9}\right)$
all terms in (\ref{eq:RT from Zhat for M(-2,1/2,1/3,1/2)}) are nontrivial.

Equation (\ref{eq:RT from ST}) leads to
\begin{align}
\textrm{RT}\left[M\left(-2;\frac{1}{2},\frac{1}{3},\frac{1}{2}\right);\frac{r}{s}\right]= & \frac{F\left[\Gamma_{M\left(-2;\frac{1}{2},\frac{1}{3},\frac{1}{2}\right)};\frac{r}{s}\right]}{F[-1\bullet;\frac{r}{s}]^{4}},\label{eq:RT from ST for M(-2,1/2,1/3,1/2)}\\
F\left[\Gamma_{M\left(-2;\frac{1}{2},\frac{1}{3},\frac{1}{2}\right)};\frac{r}{s}\right]= & \left.\frac{\sum_{n\in\{1,\ldots,s-1\}^{4}}\prod_{v\in\textrm{V}}T_{n_{v}n_{v}}^{a_{v}}S_{0n_{v}}^{2-\deg(v)}\prod_{(v_{1},v_{2})\in E}S_{n_{v_{1}}n_{v_{2}}}}{\left(q^{1/2}-q^{-1/2}\right)^{5}}\right|_{q=e^{2\pi i\frac{r}{s}}},\nonumber 
\end{align}
where $S$ and $T$ matrices are the same as in (\ref{eq:S =000026 T matrices Sigma(2,3,7)})
except 
\begin{equation}
a_{v}=\begin{cases}
-2 & v=1,2,4\\
-3 & v=3
\end{cases}
\end{equation}
Using Mathematica we checked that (\ref{eq:RT from Zhat for M(-2,1/2,1/3,1/2)})
and (\ref{eq:RT from ST for M(-2,1/2,1/3,1/2)}) give the same result.
Similarly to $M\left(-1;\frac{1}{2},\frac{1}{3},\frac{1}{9}\right)$
the necessity of calculating $\textrm{Coker}(rM)$ for each $r$ made
it easier to increase the parameter $s$ (however in this case the
cokernel is bigger) and we stopped at $r/s=5/21$.

\section{Open questions \label{sec:Open-questions}}

The most interesting future direction seems to be the one towards
the interpretation of our main conjecture. Do we really have another
manifold associated to each $r$? The manifold corresponding to the
matrix $rM$ is not an $r$-fold cover of the one corresponding to
$M$ and it is difficult to find another topologically reasonable
candidate. Or maybe the interpretation should not involve another
manifold? But what would the summation over $\textrm{Coker}(rM)$
mean in this case?

Another goals for future research are the proof of our main conjecture
and an investigation of 3-manifolds that are not Seifert and -- more
generally -- not plumbed.

\section*{Acknowledgements}

I would like to thank Sergei Gukov for his mentoring, Sungbong Chun
for invaluable support, and Miranda Cheng for directing my attention
to this topic. I am also grateful to Francesca Ferrari, Sarah Harrison,
Thang Le, Pavel Putrov, and Piotr Su\l kowski for insightful discussions.
My work is supported by the Polish Ministry of Science and Higher
Education through its programme Mobility Plus (decision no. 1667/MOB/V/2017/0).
This research was supported in part by the National Science Foundation
under Grant No. NSF PHY-1748958 while I was visiting Kavli Institute
for Theoretical Physics in Santa Barbara.

\bibliographystyle{JHEP}
\bibliography{Kuch1906}

\end{document}